# Origin of Stripe and Quasi-Stripe CDW Structures in Monolayer MX$_2$ compounds: Multivalley Free Energy Landscape

Keiji Nakatsugawa[1,2*†], Satoshi Tanda[1,2*‡] and Tatsuhiko N. Ikeda[3*§]

[1]*Department of Applied Physics, Hokkaido University, Kita 13, Nishi 8, Kita-ku, Sapporo, 0608628, Hokkaido, Japan. Tel.: +81-11-706-6154, Fax: +81-11-706-6154*

[2]*Center of Education and Research for Topological Science and Technology, Hokkaido University, Kita 13, Nishi 8, Kita-ku, Sapporo, 060-8628, Hokkaido, Japan. Tel.: +81-11-706-6154, Fax: +81-11-706-6154*

[3]*Institute for Solid State Physics, University of Tokyo, Kashiwa, 277-8581, Chiba, Japan*

**Ultrathin sheets of transition metal dichalcogenides (MX$_2$) with charge density waves (CDWs) is increasingly gaining interest as a promising candidate for graphene-like devices. Although experimental data including stripe/quasi-stripe structure and hidden states have been reported, the ground state of ultrathin MX$_2$ compounds and, in particular, the origin of anisotropic (stripe and quasi-stripe) CDW phases is a long-standing problem. Anisotropic CDW phases have been explained by Coulomb interaction between domain walls and inter-layer interaction. However, these models *assume* that anisotropic domain walls can exist in the first place. Here, we report that anisotropic CDW domain walls can appear naturally without assuming anisotropic interactions: We explain the origin of these phases by topological defect theory (line defects in a two-dimensional plane) and interference between harmonics of macroscopic CDW wave**

---
[*] Corresponding author.
[†] E-mail: keiji_nakatsugawa@eis.hokudai.ac.jp
[‡] E-mail: tanda@eng.hokudai.ac.jp
[§] E-mail: tikeda@issp.u-tokyo.ac.jp

**functions. We revisit the McMillan-Nakanishi-Shiba model for monolayer 1*T*-TaS$_2$ and 2*H*-TaSe$_2$ and show that CDWs with wave vectors that are separated by 120° (i.e. the three-fold rotation symmetry of the underlying lattice) contain a free-energy landscape with many local minima. Then, we remove this 120° constraint and show that free energy local minima corresponding to the stripe and quasi-stripe phase appear. Our results imply that Coulomb interaction between domain walls and inter-layer interaction may be secondary factors for the appearance of stripe and quasi-stripe CDW phases. Furthermore, this model explains our recent experimental result (appearance of the quasi-stripe structure in monolayer 1T-TaS2) and can predict new CDW phases, hence it may become the basis to study CDW further. We anticipate our results to be a starting point for further study in two-dimensional physics, such as explanation of "Hidden CDW states", study the interplay between supersolid symmetry and lattice symmetry, and application to other van der Waals structures.**

Usually, anisotropic structures such as stripe phases can be explained by anisotropic interaction between the constituent atoms, electrons, or liquid crystal polymers [1]. Anisotropic structures in charge density waves (CDWs) have been explained likewise [2, 3, 4, 5]. CDWs are periodic modulations of electric charge density in low-dimensional conductors [6, 7, 8, 9, 10]. The stability of CDWs containing domain stripe walls [11, 12] and quasi-stripe (triclinic) domain walls [13, 14, 12] have been explained by Coulomb interaction between domain walls [2, 3] and three-dimensional stacking [4]. However, are these interactions indispensable? It would be ideal to have rich structures with the least amount of interactions.

In this article we report that stripe and quasi-stripe CDW domain walls can appear without anisotropic interactions and explain the origin of these phases by topological defect theory (line defects in a two-dimensional plane) and interference between harmonics of macroscopic CDW wave functions. We consider the transition metal dichalcogenide ($MX_2$) compounds $1T$-$TaS_2$ and $2H$-$TaSe_2$ (Figure 1 (a),(b)) which have recently experienced a resurgence of interest due to rich physical content and potential applicability to nanoscale electromechanics [15, 16, 17, 18, 19, 20, 21, 22, 23, 24] [25, 26, 27]. These materials exhibit various CDW phases which are characterized by domain walls (topological defects). Interesting CDW phases are the stripe phase with stripe domain walls and the triclinic (T) phase with quasi-stripe domain walls, which surprisingly show up only on *heating* from the C phase (Figure 1 (d) and (e)).

Our method is to use a Ginzburg-Landau model for a CDW with general wave vectors $\mathbf{Q}^{(i)}$ ($i = 1, 2, 3$) satisfying the triple-Q condition $\mathbf{Q}^{(1)} + \mathbf{Q}^{(2)} + \mathbf{Q}^{(3)} = \mathbf{0}$ (see Methods for detail). Historically, this model was first introduced by McMillan [28, 29, 2] to explain the IC-C phase transition. Nakanishi and Shiba analyzed this model more carefully by including higher order harmonics in the order parameter configuration, finding the presence of the NC phase in $1T$-$TaS_2$ [30] (Figure 1 (d)). Experimentally, the C, IC, NC, stripe and T phases all satisfy the triple-Q condition. It is noteworthy that the free energy only involves the terms compatible with the crystal symmetry and the only input from experimental data is the incommensurate wave vectors $\mathbf{Q}_{IC}^{(i)}$ for the IC phase. Nevertheless, as shown by Nakanishi and Shiba, the free energy has an unexpected minimum corresponding to the NC phase. Thus, it is natural to ask whether there are more hidden minima if one makes a thorough search which was not feasible at the time.

We revisit the McMillan-Nakanishi-Shiba models [2, 30, 31] for monolayer 1$T$-TaS$_2$ and 2$H$-TaSe$_2$ without interaction between domain walls. First, we show that CDWs with $\mathbf{Q}^{(i)}$s that are separated by 120° (i.e. the three-fold rotation symmetry of the underlying lattice) contain a free-energy landscape with many local minima. Then, we remove this constraint and show that free energy local minima corresponding to the stripe and T phases appear. Finally, we explain the origin of stripe and T domain walls and discuss the implication of our results.

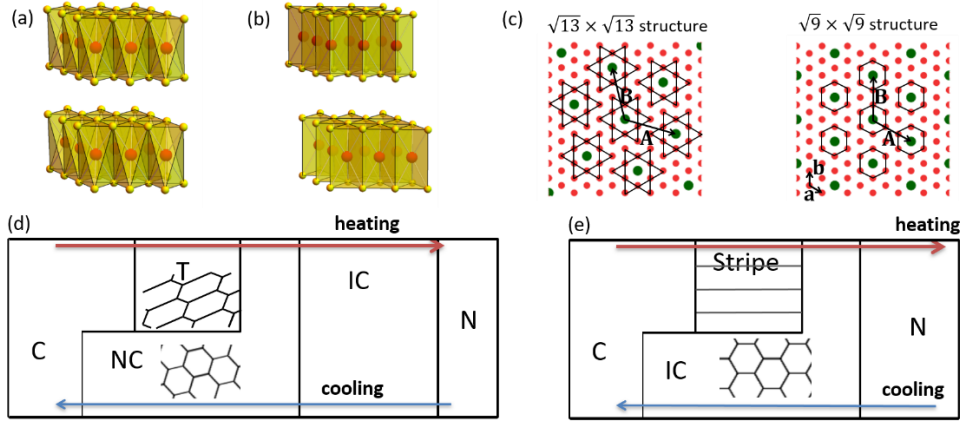

Figure 1: CDW structures in 1$T$-TaS$_2$ and 2$H$-TaSe$_2$. (a): Structure of 1$T$-TaS$_2$. (b): Structure of 2$H$-TaSe$_2$. These materials have two-dimensional layered crystalline structures like graphene. Here, the red spheres represent Ta atoms and the yellow spheres represent S or Se atoms. (c):The Ta atoms form a triangular lattice with superlattice vectors $\mathbf{a}$ and $\mathbf{b}$. $\mathbf{A}$ and $\mathbf{B}$ are commensurate (C) CDW superlattice vectors ($\mathbf{A} = (\mu + \nu)\mathbf{a} + \nu\mathbf{b}$, $\mathbf{B} = -\nu\mathbf{a} + \mu\mathbf{b}$, $|\mathbf{A}| = |\mathbf{B}| = \sqrt{\mu^2 + \mu\nu + \nu^2}|\mathbf{a}|$). The CCDW superlattice for 1$T$-TaS$_2$ ($\mu = 3, \nu = 1$) and 2$H$-TaSe$_2$ ($\mu = 3, \nu = 0$) are shown with green spheres. (d) Temperature dependence of CDW phases in bulk 1$T$-TaS$_2$. (e) Temperature dependence of CDW phases in bulk 2$H$-TaSe$_2$. The blue arrows represent the cooling cycle. The red arrows represent the heating cycle. C: Commensurate phase, T: Triclinic phase with quasi-stripe domain walls (stretched honeycomb lattice). NC: Nearly commensurate phase, IC: Incommensurate phase, N: Normal metal phase. The domain walls of the NC, T, and stripe phases are depicted with solid lines.

# 1 Results

## 1.1 1$T$-TaS$_2$ CDW States with 120° Constraint

First, we consider the special case that the CDW wave vectors $\mathbf{Q}^{(i)}$ ($i = 1,2,3$) are separated by 120°.

In this case, the triple-Q condition $\mathbf{Q}^{(1)} + \mathbf{Q}^{(2)} + \mathbf{Q}^{(3)} = \mathbf{0}$ implies $|\mathbf{Q}^{(1)}| = |\mathbf{Q}^{(2)}| = |\mathbf{Q}^{(3)}|$. Numerical results of CDW free energy $F[\{\mathbf{Q}^{(i)}\}; T]$ at temperature $T$ are shown in Figure 2. Each $\mathbf{Q}^{(i)}$s ($i = 1,2,3$) define a two-dimensional reciprocal space. To characterize the set $\{\mathbf{Q}^{(i)}\}$ we use two components of $\mathbf{Q}^{(1)}$ since they uniquely determine $\mathbf{Q}^{(2)}$ and $\mathbf{Q}^{(3)}$ by the triple-Q condition and the 120° constraint. The CDW free energy can be visualized by varying these components.

Figure 2 (a) reproduces the results by Nakanishi-Shiba which uses the special wave vector path $\mathbf{Q}^{(1)} = \mathbf{Q}(x)$ (shown in Figure 2 (b) with black lines). We have used the free-energy parameters in ref. [30]. Important wave vectors are the commensurate wave vectors $\mathbf{Q}_C^{(i)}$ and the incommensurate wave vectors $\mathbf{Q}_{IC}^{(i)}$ which have the norms $|\mathbf{Q}_C^{(i)}|/|\mathbf{G}_i| = 1/\sqrt{13} \approx 0.277$ and $|\mathbf{Q}_{IC}^{(i)}|/|\mathbf{G}_i| = 0.283$, respectively, and are tilted from the primitive reciprocal lattice vectors $\mathbf{G}_i$ by angles of 13.9° and 0°, respectively. $N = 0$ represents the charge density modulation by the fundamental wave $\mathbf{Q}^{(i)}$. The important idea by Nakanishi-Shiba is to include higher order harmonics with $N = 1,2,3, ...$ (see Figure 6). In fact, the NC phase appears for $N > 0$.

In our extended analysis we have used two types of free energies: the first one (type-1) uses phenomenological parameters which reproduce a ring-like diffuse scattering obtained by an electron diffraction experiment [32], while the second one (type-2) uses McMillan's original free energy [2] analyzed with the method of Nakanishi-Shiba (see the Method section for more detail). The type-2 free energy is important because ring-like diffuse scattering were not observed for ultra-thin sheet of 1$T$-TaS$_2$ including monolayer [22]. Figure 2 (b), (c), (d) show the free energy in the $\mathbf{Q}^{(1)}$ space. Note that the same free energy is obtained for $\mathbf{Q}^{(2)}$ and $\mathbf{Q}^{(3)}$ because of the three-fold rotational symmetry.

Surprisingly, the free energy has a "multivalley landscape": there are many local minima besides those corresponding to the well-known IC, NC, and C CDWs. Each of these local minima correspond to new CDW states with three-fold rotational symmetry. Despite of different global texture, type-1 and type-2 free energies show almost identical local minima. If the 120° constraint is removed, then these local minima are expected to move to new minima corresponding to the T phase.

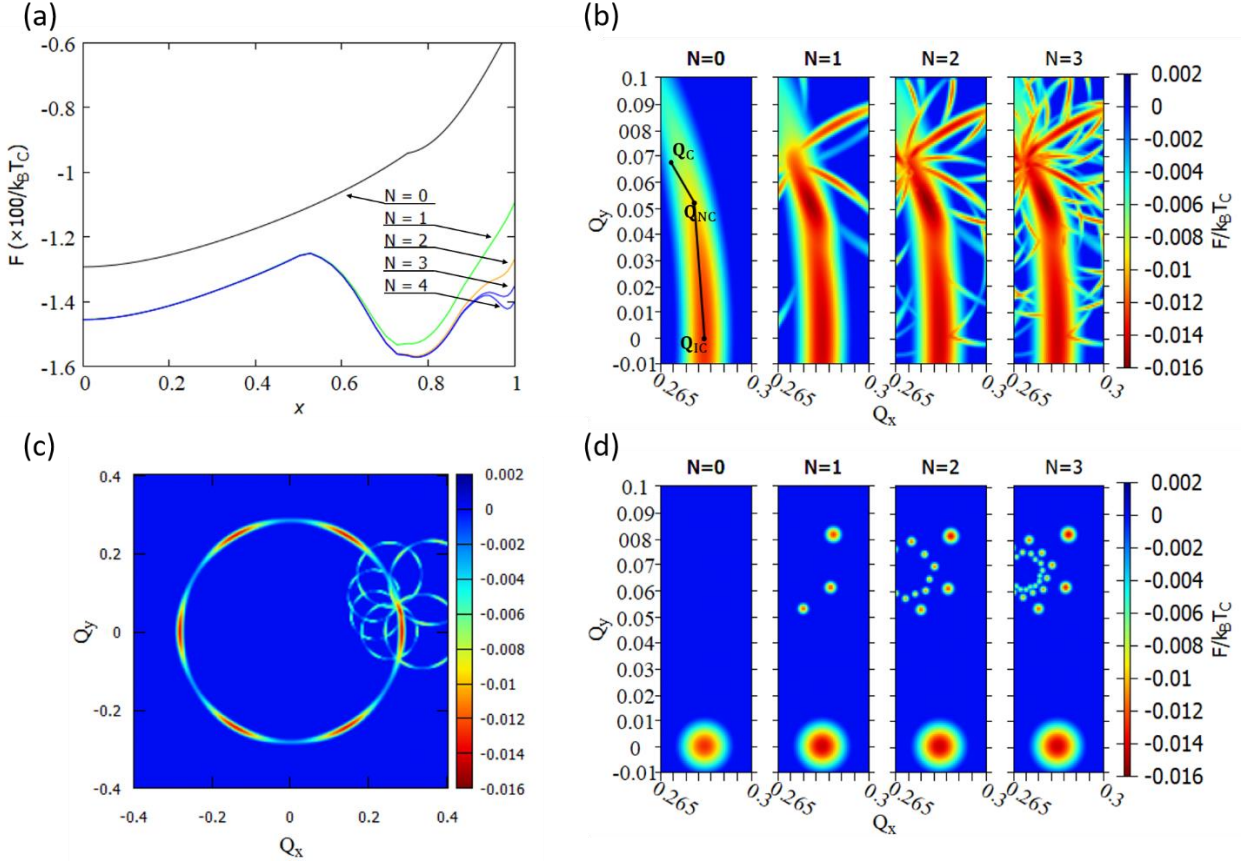

Figure 2: Numerical results of $1T$-TaS$_2$ free energy in the $\mathbf{Q}^{(1)} = (Q_x, Q_y)$ space (in units with $|\mathbf{G}_i| = 1$). (a) Reproduction of Nakanishi-Shiba's result [30]. $x$ parameterizes the wave vector $\mathbf{Q}(x)$ with $\mathbf{Q}(0) = \mathbf{Q}_{\text{IC}}$, $\mathbf{Q}(x^*) = \mathbf{Q}_{\text{NC}}$ (nearly commensurate), and $\mathbf{Q}(1) = \mathbf{Q}_{\text{C}}$. $\mathbf{Q}(x)$ is shown with solid black lines in (b). The "multivalley landscape" of the type-1 free energy with $N = 0$, $N = 1$, $N = 2$, and $N = 3$ are given in (b). (c) show the type-1 free energy for larger $Q_x, Q_y$ values. Note that the number of branches and local minima increase as $N$ increases. New local minima appear near $\mathbf{Q}_{\text{C}}$. (d) The type-2 free energy with $N = 0$, $N = 1$, $N = 2$, and $N = 3$ show almost identical local minima, but it is easier to see how the CCDW states is obtained as $N \to \infty$.

## 1.2 2H-TaSe$_2$ CDW states with 120° Constraint

The multivalley free energy structure also exists in $2H$-TaSe$_2$ as shown in Figure 3. The commensurate wave vectors $\mathbf{Q}_C^{(i)}$ and the incommensurate wave vectors $\mathbf{Q}_{IC}^{(i)}$ have the norms $|\mathbf{Q}_C^{(i)}|/|\mathbf{G}_i| = 1/\sqrt{9} \approx$ 0.333 and $|\mathbf{Q}_{IC}^{(i)}|/|\mathbf{G}_i| = 0.325$, respectively, and they are parallel to the primitive reciprocal lattice vectors $\mathbf{G}_i$. Figure 3 (a) reproduces the free energy along the special line $Q_x = 0$ obtained by Nakanishi-Shiba [31] which explains the IC-C phase transition. We extend their analysis to the two-dimensional plane by varying $Q_x$ and find new local free energy minima like those of $1T$-TaS$_2$. Here, we have used the free-energy parameters in ref. [31]. These local minima correspond to CDW states with three-fold rotational symmetry. Despite different global texture, type-1 and type-2 free energies show almost identical local minima. If the 120° constraint is removed, then these local minima are expected to move to new minima corresponding to the stripe phase.

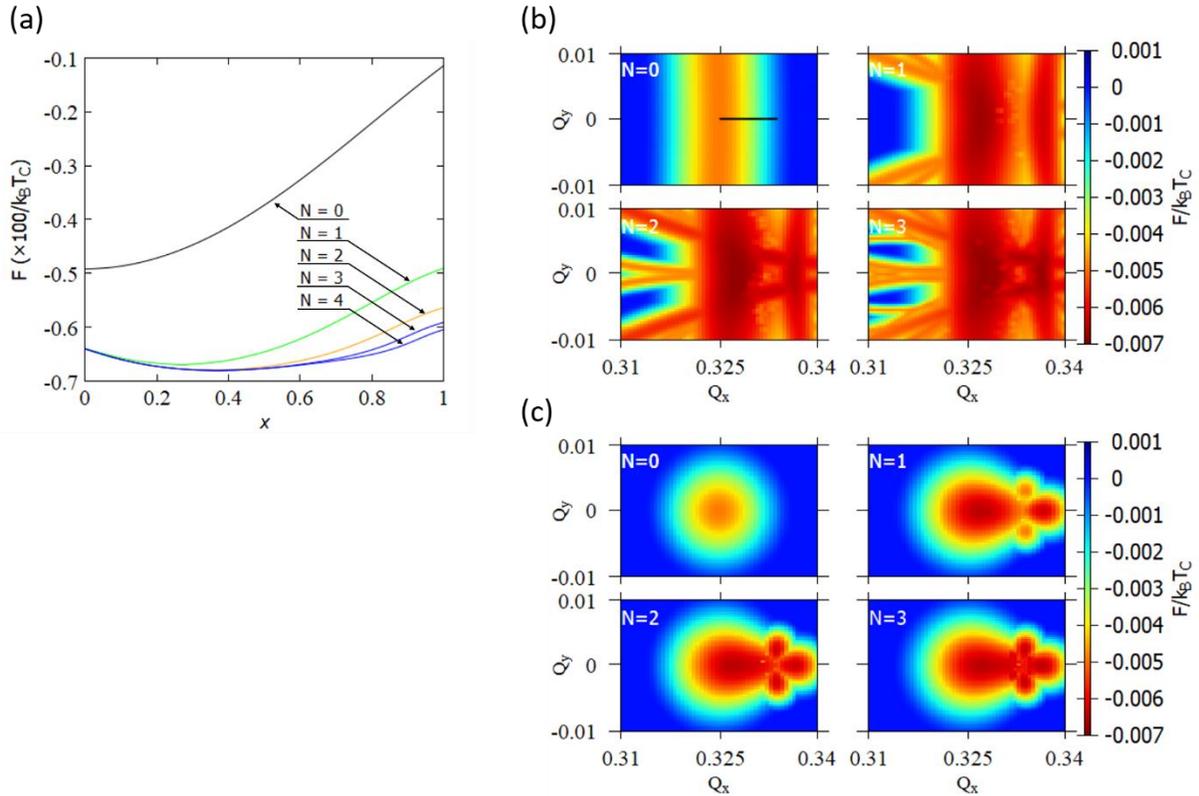

Figure 3. Numerical results of $2H$-TaSe$_2$ free energy in the $\mathbf{Q}^{(1)} = (Q_x, Q_y)$ space (in units with $|\mathbf{G}_i| = 1$). (a): Reproduction of the result by Nakanishi-Shiba [31]. $x$ parameterizes the wave vector $\mathbf{Q}(x)$ with $\mathbf{Q}(0) = \mathbf{Q}_{IC}$, $\mathbf{Q}(x^*) = \mathbf{Q}_{NC}$ (nearly commensurate), and $\mathbf{Q}(1) = \mathbf{Q}_C$. $\mathbf{Q}(x)$ is shown with a black line in (b) The "multivalley

landscape" of the free energy with $N = 0$, $N = 1$, $N = 2$ and $N = 3$ are given for the type-1 (b) and type-2 (c) free energies.

## 1.3: Anisotropic CDW Phases

Here, we consider CDW states without the 120° constraint. Let $R_\pm$ denote the matrices which rotate a wave vector $\mathbf{Q} = (Q_x, Q_y)$ by $\pm 120°$. If we set the 120° constraint, then the vectors $\{\mathbf{Q}^{(1)}, R_-\mathbf{Q}^{(2)}, R_+\mathbf{Q}^{(3)}\}$ are identical. If the 120° constraint is removed, then the triple-Q condition implies that the vertices of $\{\mathbf{Q}^{(1)}, R_-\mathbf{Q}^{(2)}, R_+\mathbf{Q}^{(3)}\}$ form a regular triangle (see Figure 6). Consequently, stable or quasi-stable CDW states with anisotropic domain walls are obtained if local minima from the free energy shown with $\mathbf{Q}^{(1)}$, $\mathbf{Q}^{(2)}$, $\mathbf{Q}^{(3)}$ form a regular triangle.

First, we consider the T phase. From the triple-Q condition there are four independent degrees of freedom. Here, we fix two degrees of freedom, namely, the angles between domain walls, to visualize the free energy. The angles between domain walls are known from experiments [14]. Here, we consider the angles $\delta\phi_1 = 360° - \delta\phi_2 - \delta\phi_3$, $\delta\phi_2 = 180° - \phi_C$, $\delta\phi_3 = 150°$, where $\phi_C \approx 13.9°$ is the angle between $\mathbf{Q}_C^{(i)}$ and $\mathbf{Q}_{IC}^{(i)}$. Figure 4 (a) shows the free energy with $N = 1$. We have used the free energy parameters from Nakanishi-Shiba [30]. Then, we find a triplet of local minima which forms a regular triangle (black). This result is surprising because the T phase was previously explained by inter-layer interaction [4], but our calculation is done for a monolayer crystal. Experimental result of the T phase in bulk $1T$-TaS$_2$ (green triangle) [14] is in good agreement with our calculation. The domain size in our calculation is smaller than bulk crystal but may become closer to experiment if we further consider Coulomb interaction between domain walls or inter-layer interaction. Aside from this quantitative detail, we conclude that anisotropic domain walls for the T phase can be formed without inter-layer interaction and, hence, in a monolayer $1T$-TaS$_2$.

Next, we consider the stripe phase. According to experimental results [11], the stripe phase is obtained by minimizing the free energy with the constraint $\mathbf{Q}^{(1)} = \mathbf{Q}_C^{(1)}$. Figure 4 (b) is the free energy ($N = 1$) visualized with $\mathbf{Q}^{(2)}$ and $\mathbf{Q}^{(3)}$. Clearly, these local minima form a regular triangle with $\mathbf{Q}_C^{(1)}$. These domain walls correspond to the stripe phase. Again, the domain size is smaller than the stripe domain of bulk $2H$-TaSe$_2$ (green triangle) [11]. The domain size may become closer to experiment if we further consider Coulomb interaction between domain walls. Therefore, we conclude that anisotropic domain walls for the stripe phase can be formed naturally even in a monolayer $2H$-TaSe$_2$ with the least amount of interaction.

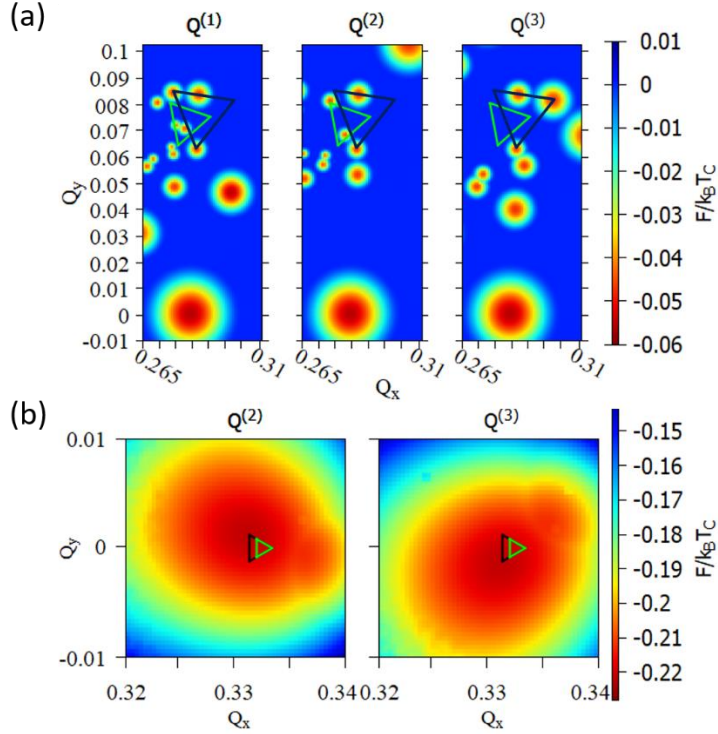

Figure 4. Free energy of T and stripe phases. (a) Free energy of $1T$-TaS$_2$ without the 120° constraint ($T = 225K$). Local minima for $\mathbf{Q}^{(1)}$, $\mathbf{Q}^{(2)}$ and $\mathbf{Q}^{(3)}$ form a regular triangle. These local minima correspond to the T phase with quasi-stripe domain walls. (b) Free energy of $2H$-TaSe$_2$ without the 120° constraint ($T = 100K$). The point $\mathbf{Q}^{(1)} = \mathbf{Q}_C^{(1)}$ and local minima for $\mathbf{Q}^{(2)}$ and $\mathbf{Q}^{(3)}$ form a regular triangle. Therefore, these local minima correspond to the stripe phase.

## 2   Discussion and Conclusion

First, we summarize the results presented in the previous sections. We revisited the McMillan-Nakanishi-Shiba models for $1T$-TaS$_2$ and $2H$-TaSe$_2$ with general CDW wave vectors $\mathbf{Q}^{(i)}$ ($i = 1,2,3$) separated by 120°. A "bird's-eye view" of CDW free energy landscape with multivalley structure reveal the presence of multiple local minima which correspond to different CDW states. Then, we removed the 120° constraint and observed local minima for the stripe and T phases. These results are surprising because we resorted to neither Coulomb interaction between domain walls nor inter-layer interaction, which had been considered as the origins of the stripe and phases.

To explain these results, we consider the following mechanism. CDW phases between the IC and C phases are accompanied with domain walls, where each domain contains a CCDW with a different phase: such domain walls minimize the commensurability energy between the CDW and the underlying lattice. The average domain size is inversely proportional to $|\mathbf{Q}^{(i)} - \mathbf{Q}_C^{(i)}|$. As temperature decreases, $\mathbf{Q}^{(i)}$ approaches $\mathbf{Q}_C^{(i)}$ and the domain size diverges: The ground state is the C phase without domain walls. As temperature increases from the ground state, energy increase due to thermal fluctuation induces domain walls. Since the formation of domain walls requires energy, these domain walls are created one by one. These domain walls are expected to be anisotropic rather than hexagonal, since the formation of hexagonal domain walls require a global arrangement. This model is analogous to the formation of the Abrikosov vortex lattice in a type-2 superconductor, where vortices are formed one by one and finally form a lattice structure as the number of vortices increase.

Moreover, for 1T-TaS$_2$, the C phase is closer (in the reciprocal lattice space) to the T phase than the NC phase, i.e. the T phase has larger domains which is energetically favored. Therefore, as temperature increases, 1T-TaS$_2$ will first make a phase transition from the C phase to the T phase; phase transition to the NC phase occurs when the NC domain size and the T domain size become comparable (Figure 5) [14].

We note that this induced anisotropy model has experimental support. The low temperature CDW phase of a free-standing monolayer 1*T*-TaS$_2$ with multiple T domain walls has been observed with scanning transmission electron microscopy [22]. In addition, 1*T*-TaS$_2$ with a *single* domain wall has been observed using scanning tunneling microscopy [33].

Next, we explain what condition distinguishes stripe and T domain walls. While 2*H*-TaSe$_2$ has a parity-conserving commensurate structure (Figure 1(c)), 1*T*-TaS$_2$ breaks parity in the first place, and hence they cannot have stripe domain walls. So, from our results we predict that MX$_2$ compounds with triple-CDW can have domain walls with an angle of $180° - \phi_C$, where $\phi_C$ is the angle between $\mathbf{Q}_C^{(i)}$ and $\mathbf{Q}_{IC}^{(i)}$. From the commensurability condition conditions $\mu \mathbf{Q}_C^{(i)} - \nu \mathbf{Q}_C^{(i+1)} = \mathbf{G}_i$ where $\mathbf{G}_i$ are reciprocal lattices of the crystal, we obtain $\phi_C = \cos^{-1}\frac{\mu+\nu/2}{\sqrt{\mu^2+\mu\nu+\nu^2}}$. Then, we see that $\mu \neq 0$ and $\nu \neq 0$ gives T domain walls (such as 1$T$-TaS$_2$ with $\mu = 3$ and $\nu = 1$), while $\mu \neq 0$ and $\nu = 0$ gives stripe domain walls (such as 2$H$-TaSe$_2$ with $\mu = 3$ and $\nu = 0$). In particular, we predict that the ultrathin TaSe$_2$ with a new $\sqrt{7} \times \sqrt{7}$ commensurate structure (that is, $\mu = 2$ and $\nu = 1$) [34] can also have T domain walls.

The appearance of the stripe and T phases with the least amount of interaction can also be explained in the context of entropy. The Kosterlitz-Thouless transition [35, 36], for instance, is stabilized by the production of vortices. Creation of vortices requires energy, but vortices also increase entropy and minimize the free energy of the system. Similarly, stabilization of the stripe and T phases is related to entropy release/increase which is asymmetric with heating/cooling. The stripe phase and the T phase break the 3-fold rotational symmetry of the CCDW state. The T phase further breaks parity. Lower symmetry due to production of symmetry-breaking domain walls implies larger entropy, i.e. lower free energy.

From these discussions, we conclude that the T and stripe phases can appear naturally even in monolayer MX$_2$. Our results imply that Coulomb interaction between domain walls or inter-layer interaction may be secondary factors for the appearance of these phases. Note that our result does not contradict with Nakanishi-Shiba's model of stripe and T phases based on inter-layer stacking [4]: their work still apply for MX$_2$ with higher number of layers. Our result also explains our recent experimental result, namely the appearance of the T phase in *monolayer* 1$T$-TaS$_2$.

As future applications of this work, the appearance of various local minima may explain mysterious CDW states such as the "hidden CDW state" [37, 38, 39, 24] and these minima also predict new CDW phases, hence our model may become the basis to study CDW further. Our analysis may be extended to study the interplay between supersolid symmetry and lattice symmetry, and application to other van der Waals structures [40].

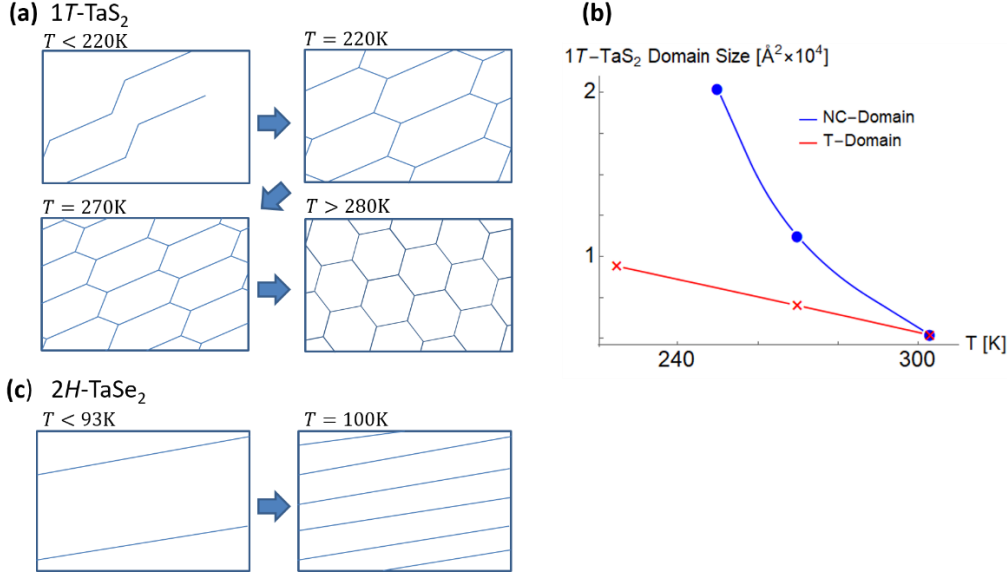

Figure 5: A model for the formation of anisotropic domain walls. (a): The ground state of $1T$-TaS$_2$ is the C phase without domain walls. As temperature increases, domain walls are formed one by one. As the number of domain walls increase, we obtain the T phase. The domain size decreases as temperature increase further. Then, T-NC phase transition occur when the NC domain size and the T domain size become comparable ($280K$ for bulk $1T$-TaS$_2$). The temperature dependence is shown here for bulk $1T$-TaS$_2$ [14]. (b) Bulk data of $1T$-TaS$_2$ domain size. The T domain size (heating) are from ref. [14]. The NC domain size (cooling) are from ref. [41]. We note that the same NC domain sizes were verified but not reported in ref. [14]. (c) Similarly, the ground state of $2H$-TaSe$_2$ is the CCDW state without domain walls. As temperature increases, we speculate that domain walls are formed one by one. The temperature dependence is shown here for bulk $2H$-TaSe$_2$ [11].

## 3  Methods

McMillan's phenomenological free energy for triple CDW is [28, 29, 2]

$$F = \int d^2r \{a(\mathbf{r})\alpha^2(\mathbf{r}) - b(\mathbf{r})\alpha^3(\mathbf{r}) + c(\mathbf{r})\alpha^4(\mathbf{r}) + \Sigma_{i=1}^3 \psi_i^*(\mathbf{r})e(-i\nabla)\psi_i(\mathbf{r})$$

$$+ d(\mathbf{r})[|\psi_1(\mathbf{r})\psi_2(\mathbf{r})|^2 + |\psi_2(\mathbf{r})\psi_3(\mathbf{r})|^2 + |\psi_3(\mathbf{r})\psi_1(\mathbf{r})|^2]\}$$

Here the complex order parameters $\psi_i(\mathbf{r})$ at position $\mathbf{r}$ are related to the electron charge density

$\rho(\mathbf{r}) = \rho_0(\mathbf{r})\{1 + \alpha(\mathbf{r})\}, \alpha(\mathbf{r}) = \sum_{i=1}^{3} \text{Re}[\psi_i(\mathbf{r})]$. $\rho_0(\mathbf{r})$ is the charge density in the normal state. The index $i\ (=1,2,3)$ specifies the three components of the triple CDW. $a(\mathbf{r}), b(\mathbf{r}), c(\mathbf{r}), d(\mathbf{r})$ are position-dependent coefficients which have the periodicity of the crystal lattice and can be written in the form $a_0 + a_1 \sum_{i=1}^{6} e^{i\mathbf{G}_i \cdot \mathbf{r}}$, etc., where $\mathbf{G}_i$ the six shortest reciprocal lattice vectors. The commensurate wave vectors $\mathbf{Q}_C^{(i)} = |\mathbf{Q}_C^{(i)}|(\cos\phi_{Ci}, \sin\phi_{Ci})$ are defined by the conditions $\mu\mathbf{Q}_C^{(i)} - \nu\mathbf{Q}_C^{(i+1)} = \mathbf{G}_i$ $(i = 1,2,3)$, $|\mathbf{Q}_C^{(i)}| = |\mathbf{G}_i|/\sqrt{\mu^2 + \mu\nu + \nu^2}$. Type-1 and Type-2 free energies explained in the Result section correspond to different forms of $e(-i\nabla)$. In their study of $1T$ and $2H$-polytypes, Nakanishi and Shiba considered $(\mu, \nu) = (3,1)$ and $(\mu, \nu) = (3,0)$, respectively, but their method can be generalized and applied to other CDW materials with different $(\mu, \nu)$ values. The complex order parameters of a CCDW state is $\psi_i(\mathbf{r}) = \Delta_C e^{i\mathbf{Q}_C^{(i)} \cdot \mathbf{r} + i\theta_{Ci}}$ where $\mathbf{Q}_C^{(i)}$ are the wave vectors of the CCDW state and $\theta_{Ci}$ are constant phases which minimize the free energy. For other CDW states, $\psi_i(\mathbf{r})$ can be expanded as a function of the deviation vectors $\mathbf{q}^{(i)} = \mathbf{Q}^{(i)} - \mathbf{Q}_C^{(i)}$. Then we may expand $\psi_i(\mathbf{r})$ as $\psi_i(\mathbf{r}) = \sum_{\substack{l,m,n\geq 0 \\ l \cdot m \cdot n = 0}} \Delta_{lmn}^{(i)} e^{i\mathbf{Q}_{lmn}^{(i)} \cdot \mathbf{r}}$ where $l, m, n$ are integers and $\mathbf{Q}_{lmn}^{(i)} = l\mathbf{k}^{(i)} + m\mathbf{k}^{(i+1)} + n\mathbf{k}^{(i+2)} + \mathbf{Q}^{(i)}$, $\mathbf{k}^{(i)} = \mu\mathbf{q}^{(i)} - \nu\mathbf{q}^{(i+1)}$ and $\mathbf{q}^{(i)} = \mathbf{Q}^{(i)} - \mathbf{Q}_C^{(i)}$, and we have introduced the cyclic notations such as $\mathbf{q}^{(4)} = \mathbf{q}^{(1)}$ and $\mathbf{q}^{(5)} = \mathbf{q}^{(2)}$. The free-energy is obtained by varying general $\mathbf{Q}^{(i)}$s. In this case, we assume that each $\Delta_{lmn}^{(i)}$ are real. These wave vectors are depicted in Figure 6. $\Delta_{lmn}^{(i)}$ are determined by numerically calculating the set of coupled non-linear differential equations $\frac{\partial F}{\partial \Delta_{lmn}^{(i)}} = 0$.

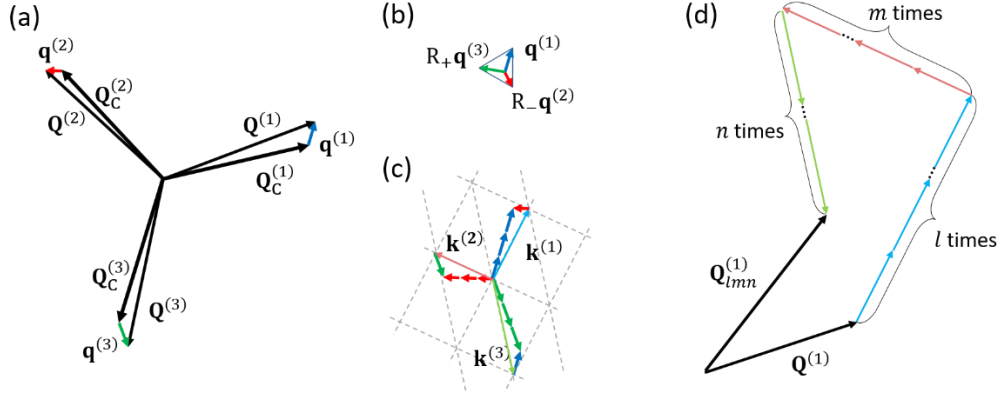

Figure 6: The elements of the McMillan-Nakanishi-Shiba model. Note that in this article we focus on $|\mathbf{q}^{(1)}| = |\mathbf{q}^{(2)}| = |\mathbf{q}^{(3)}|$ and each $|\mathbf{q}^{(i)}|$ are separated by a 120°. (a) CCDW wave vectors $\mathbf{Q}_C^{(i)}$ define the CCDW charge density $\rho(\mathbf{r}) = \rho_0[1 + \Delta_C \sum_{i=1}^{3} \cos(\mathbf{Q}_C^{(i)} \cdot \mathbf{r} + \theta_{C\,i})]$. The free energy is calculated as a function of a general wave vector $\mathbf{Q}^{(i)}$ with the triple-$\mathbf{Q}$ condition $\mathbf{Q}^{(1)} + \mathbf{Q}^{(2)} + \mathbf{Q}^{(3)} = 0$. (b) The deviation vectors $\mathbf{q}^{(i)} = \mathbf{Q}^{(i)} - \mathbf{Q}_C^{(i)}$ projected on the $\mathbf{Q}^{(1)}$ space form a regular triangle. Here, $R_\pm$ are rotation matrix of $\pm 120°$. If $\mathbf{Q}^{(i)}$ are separated by 120°, then this triangle reduces to a point. (c) The harmonics form a triangular lattice. $\mathbf{k}^{(i)}$ are wave vectors responsible for the formation of domain walls (the image shows an example for $(\mu, \nu) = (3,1)$). (d) A general CDW state is given by $\rho(\mathbf{r}) = \rho_0[1 + \sum_{i=1}^{3} \sum_{\substack{l,m,n \geq 0 \\ l \cdot m \cdot n = 0}} \Delta_{lmn}^{(i)} \cos(\mathbf{Q}_{lmn}^{(i)} \cdot \mathbf{r})]$ with $3N(N+1)$ higher harmonics.

## Acknowledgements

We thank D. Mihailovic, M. Kosterlitz, D. Nelson, and K. Inagaki for stimulating discussions. T.N.I. acknowledge support by JSPS KAKENHI Grant No. JP18K13495.

## Authors' contributions

All authors have contributed equally.

## Competing interests

The authors declare no competing interests.